\documentstyle[12pt,fleqn,epsf]{article}
\def\ap#1#2#3{     {\it Ann. Phys. (NY) }{\bf{#1},} (19#3) #2 }

\def\jpa#1#2#3{   {\it J. Phys.} { A \bf{#1},} (#3) #2 }

\def\pla#1#2#3{    {\it Phys. Lett. }{ A \bf{#1},} (19#3) #2 }
\def\prd#1#2#3{    {\it Phys. Rev. }{ D \bf{#1},} (19#3) #2 }
\def\pra#1#2#3{    {\it Phys. Rev. }{ A \bf{#1},} (#3) #2 }

\def\prl#1#2#3{    {\it Phys. Rev. Lett. }{\bf{#1},} (19#3) #2 }

\def\ibid#1#2#3{   {\it ibid. }{\bf{#1},} (#3) #2 }

\def\eq#1{{eq.~(\ref{#1})}}

\newcommand{\mc}{\multicolumn}
\newcommand{\bea}{\begin{eqnarray}}
\newcommand{\beq}{\begin{equation}}
\newcommand{\eea}{\end{eqnarray}}
\newcommand{\eeq}{\end{equation}}
\newcommand{\nnu}{\nonumber}

\begin{document}
\title{\bf Spectrum of One-Dimensional Anharmonic Oscillators }
{\small
\author{H. A. Alhendi$^1$\ and \ E. I. Lashin$^{1,2}$\\
$^1$ Department of physics and Astronomy, College of Science,\\ King Saud University, Riyadh,
Saudi Arabia \\
$^2$ Department of Physics, Faculty of Science, \\Ain Shams University, Cairo, Egypt}
}
\maketitle
\begin{abstract}
We use a power-series expansion to calculate the eigenvalues
of anharmonic oscillators bounded by two infinite walls.
We show that for large finite values of the separation of
the walls, the calculated eigenvalues are of the same
high accuracy as the values recently obtained for the unbounded case by
the inner-product quantization method. We also apply our method
to the Morse potential. The eigenvalues obtained in this case are
in excellent agreement with the exact values
for the unbounded Morse potential.
\end{abstract}
\section{Introduction}

The spectra of many important potential functions frequently
encountered in quantum mechanics can not be obtained exactly.
Moreover, in most cases, the conventional approximate methods commonly
discussed in most standard textbooks(see, for example, \cite{landau}),
while possibly  leading to rough estimates, are either unsatisfactory or
computationally complicated. A well-known important example in
this respect is a sufficiently deep double-well potential, in which
splitting of energy levels of two neighboring states, in
particular the ground and first excited state, is extremely small
because of the non-degenerate nature of energy eigenvalues of bound states
for one-dimensional potentials \cite{landau,coleman-dw}.
Perturbation and semiclassical methods are of limited use in such a case,
and the variational method, although providing upper bounds,
requires carefully chosen trial functions. A need thus
arises for a relatively simple and effective approximate method with
a high degree of accuracy.

A variant of approximate methods and numerical techniques  has recently
been devised to calculate to high precision  the spectrum of the
one-dimensional symmetric anharmonic oscillators, with either a symmetric
solution~\cite{barakat-an}--\cite{trot-an} or a symmetry-broken
solution \cite{wang1-al}--\cite{roy-al}.

A method based on the Hill determinant has been developed in a
recent study~\cite{wang1-al}. This method, known as the inner-product
quantization procedure, uses
the representation $\Psi(x)=\sum_{i} a_i(E)\, x^i\, R_{\beta} (x)$
of the wave function, where $R_{\beta}$ is
an appropriate reference function. With a  heuristic argument based on
the Hill determinant method, it is shown that the zeros of the  coefficient
functions $a_{i}(E)$ approximate the exact bound state energies with
increasing accuracy as $i\rightarrow\infty$.
The method has been applied to several unbounded one-dimensional anharmonic
oscillators,  leading to highly accurate energy eigenvalues. In
particular, for a sufficiently deep double-well, the splitting of the
energies of the ground state energy and first excited state becomes apparent only
after $26$ significant digits.

It has already, however, been pointed out \cite{chaud-an,haut-an} that
not all eigenvalues obtained by the
Hill determinant method should be allowed, since the boundary condition
that $\Psi(x)\rightarrow 0\; \mbox{as}\, |x|\rightarrow \infty$ is
not incorporated into the method and may lead to incorrect results
for some values of the
coupling constants and for potentials in which $x=\infty$ is an
irregular singular point of the Schr\"{o}dinger equation.
In addition, the choice of the arbitrary  reference function
is important, as emphasized in~\cite{wang1-al}: in general,
it should fall off more slowly than the asymptotic form of the wave function.

The purpose of the present work is to apply the method of power-series
expansion to some  anharmonic potential functions bounded
by two infinite walls. We show that the  expansion of the wave function in
the form $\Psi(x)=\sum_{i} a_{i} x^{i}$
in the finite interval $-L<x<L$ with infinite walls at $x=\pm L$
leads, even for moderate values of $L$, to the same high accuracy
as has been achieved for the anharmonic potentials considered
in \cite{wang1-al}. The initial work in this
direction was carried out by Barakat and Rosner \cite{barakat-an},
for the case of a pure quartic
oscillator. These investigators showed that the lower-order  eigenvalues
tend rapidly to the values of the unbounded oscillator as $L$ is made large.
The method has subsequently been  employed for the bounded pure
anharmonic oscillators $x^{2k},\; \mbox{with}\;k=1,\cdots ,5,$ and for
shallow doubly anharmonic oscillators \cite{chaud-an}. The power series
in the finite range does
not require introducing reference functions and accommodates all
eigenvalues, since the boundary conditions are imposed at finite
$x$. The problem concerning the convergence of the wave functions
mentioned in~\cite{wang1-al} does not arise here.

The rest of the paper is organized as follows. In Sect.~2, we apply
the method of power-series expansion to anharmonic potentials,
specifically to a quartic, sextic, octic, dectic, duodectic and
quartic double-well. Also in Sect.~2, we consider potential
functions admitting power series-expansion, as exemplified by the
Morse potential. In Sect.~3, we explain the method for obtaining the
energy eigenvalues numerically, presenting our eigenvalue
calculations and comparing them with the values obtained by the
inner-product quantization method and with the exact values
available for the Morse potential. In Sect.~4, we present our conclusions.
\section{Power-series expansion solution}

To calculate the eigenvalues and eigenfunctions of the
anharmonic  oscillators bounded by infinitely high
potentials at $x=\pm L$, it is necessary to solve the eigenvalue equation
(we here assume units $\hbar=1\, ,\, 2\,m=1$)
\beq
\left[\frac{d^2}{d\,x^2}+E-V(x)\right]\Psi(x)=0,
 \label{seq}
 \eeq
with the boundary conditions $\Psi(\pm L)=0$. In the present work, we
consider polynomial potential functions and potentials admitting
power-series expansion. In the former  case, our
potentials $V(x)$ are of the form
\beq
V(x)=\mu^2 x^2 + g x^{2k},\hspace{2cm}
(k=2,\cdots,6)
\label{pot}
\eeq
Here the coupling
constant $g >0$ and the mass parameter $\mu^2$ takes real
values.

Using the power-series expansion
\beq
\Psi = \sum_{l=0}^{\infty} a_l x^l
\label{series}
\eeq
in \eq{seq}, we obtain the following recurrence formula for the
expansion coefficients:
\bea
a_l = \frac{g\, a_{l-2\,k-2} + \mu^2 a_{l-4}-E\,
a_{l-2}}{l\,(l-1)}, & & l \neq 0,1 \nonumber \\
a_l  =  0, & & l < 0 \nonumber\\
\label{rec}
\eea

The symmetry of \eq{seq} implies that the solutions fall
into two classes, even and odd. Even solutions can be obtained by
imposing $a_{0}~=~1, a_{1}~=~0 $, and odd
solutions can be obtained by imposing $a_{0}~=~0, a_{1}~=~1$.(We ignore
normalization in both cases.) The energy eigenvalues $E$ are obtained from
the condition $\Psi(L)=0$ for both even and odd solutions. Since
we are dealing with potentials admitting power-series expansion for
$|x|<L$, the power-series solutions of $\Psi(x)$ are convergent,
by a well-known theorem in differential equations \cite{codd}.

To illustrate the method for potential functions admitting
power-series expansion, we choose the Morse potential. For this potential,
which happens to be non symmetric, we can derive recurrence relations
in a manner similar to the symmetric case(performing a power-series
expansion for the potential and truncating it after certain power). The
resulting power-series solution depends linearly on both $a_0$ and $a_1$
and in addition depends on $E$. Extracting the coefficient of $a_0$ and
$a_1$ in the power-series solution and  calling them $f^0$ and
$f^1$, respectively, we then cast the wave function $\Psi(x,E)$ into
a form suitable for applying the boundary conditions:
\beq
\Psi(x,E)=f^0(x,E)\ a_0 + f^1(x,E)\ a_1 .
\label{nonsy1}
\eeq
The boundary conditions for $\Psi(\pm\,L,E)$ yield two
homogeneous linear equations with $a_0$ and $a_1$ as unknowns:
\bea
f^0(L,E)\ a_0 + f^1(L,E)\ a_1 &=& 0 \nnu \\
f^0(-L,E)\ a_0 + f^1(-L,E)\ a_1 &=& 0 \nnu . \\
\label{nonsy2}
\eea
For there to exist a nontrivial solution for $\{a_0,a_1\}$, the
following condition must be satisfied:
\beq
f^0(L,E)\, f^1(-L,E) - f^1(L,E)\, f^0(-L,E) =0 .
\label{nonsy3}
\eeq
This fact allows us to solve for the unknown energies $E$ and the
constants $a_0$ and $a_1$. For the case of the well-known Morse potential
\beq
V(x)=V_0 \left( 1-e^{-\lambda\, x}\right)^2,
\label{morse1}
\eeq
where $V_0$ and $\lambda$ are the depth and range parameters,
respectively,  the exact energy eigenvalues and eigenfunctions are given by
\bea
E_n &=& 2\,\lambda
 \sqrt{V_0}\left[\left(n+{1\over 2}\right)-\left(n+{1\over 2}\right)^2
{\lambda \over 2\,\sqrt{V_0}}\right] \nnu \\
\Psi_n(x) & = & N(n)\ y^{\alpha-n-{1\over 2}}\ e^{-y/2 }\
L^{2\alpha-2 n -1}_n(y) \nnu \\
\label{morse2}
\eea
where
$n$ must be positive and less than ${\sqrt{V_0}\over \lambda}-{1\over
2}$,
and where\\
$\alpha = \sqrt{{V_0\over\lambda^2}};\;\;\;\;\;
 y = 2\, \alpha\,e^{-\lambda x}.$\\
The functions $L_n^a\left(x\right)$ are the generalized Laguerre
polynomials, while the normalization constant $N(n)$ is explicitly
given by\\
$N(n) = \sqrt{{\lambda (2 \alpha - 2n -1) n! \over \Gamma(2\alpha-n)}}.$\\

To employ the method, we expand the potential in eq.~\ref{morse1} as
power series, truncating the series beyond, say, $x^{30}$:
\bea
V(x)&\approx &  V_0\,\left[ {x^2}\,{{\lambda }^2} - {x^3}\,{{\lambda }^3} +
  {\frac{7\,{x^4}\,{{\lambda }^4}}{12}} -
  {\frac{{x^5}\,{{\lambda }^5}}{4}} +
  {\frac{31\,{x^6}\,{{\lambda }^6}}{360}} -
  {\frac{{x^7}\,{{\lambda }^7}}{40}}+ \cdots \right. \nnu \\
  & & \left. +
  {\frac{536870911\,{x^{30}}\,{{\lambda }^{30}}}
    {132626429906095529318154240000000}} \right]. \nnu \\
\label{morse3}
    \eea
\section{Calculation of Energy Eigenvalues}

To obtain the energy eigenvalues, we approximate the power series
with a finite number of terms $\Psi_n(x,E)$. The boundary condition
for a specific value of $L$ is $\Psi_n(L,E)=0$. To find the
zeros of $\Psi_n(L,E)$ with respect to $E$, we first plot $\Psi_n(L,E)$
as a function of $E$.
Around each zero, we locate two nearby
points $E$ at which the function $\Psi_n(L,E)$ changes sign. We
use these two points in the initial iteration of the "bisection
method" for finding the zeros. In implementing the method,
we use the Mathematica (version 3) package, relying extensively on
its exact-number manipulation capabilities. We check the stability of
the numerical results
at  a certain degree of accuracy for a particular L by
increasing $n$ until the obtained value of $E$ stays fixed.

\begin{figure}[htbp]
\caption{\footnotesize
The even solution $\psi_e(E)$ (broken curve) and odd solution
$\psi_o(E)$ (solid curve) for the bounded quartic anharomnic
oscillator (potential $V(x)=x^2+x^4$), where $2\,L$ is the
separation of the walls and $I$ is the
number of non vanishing terms in the truncated expansion of the
wave function.}
\epsfxsize=8cm
\centerline{\epsfbox{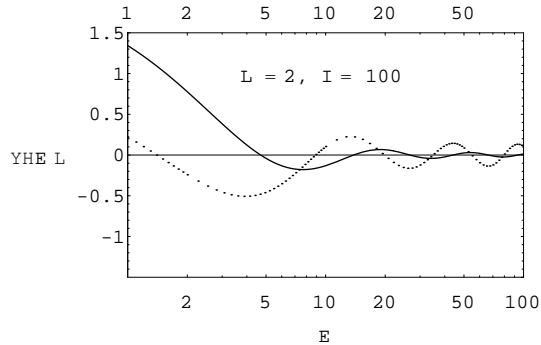}}
\label{rootconv}
\end{figure}

To illustrate the method, we present in Fig.~\ref{rootconv} an example for
the behaviour of even and odd solutions $\psi_e(E)$
and $\psi_o(E)$ for $L=2$ and $I=100$ (where $I$ indicates the number
of nonvanishing terms in the truncated expansion of the wave
function), for the case of the quartic anharmonic oscillator
(potential $V(x)=x^2+x^4$).

We use the same procedure to obtain energies in the case of a non symmetric
potential. In this case, however, we apply the condition given by
eq.~\ref{nonsy3}.

In Table~\ref{conv}, we present the calculated energies of the
ground and first excited states for the bounded quartic anharmonic
oscillator. Our method shows systematic convergence for
increasing values of both $I$ and $L$, exceeding some of the high-
accuracy results in\cite{wang1-al}. As providing evidence for the
accuracy attained, we present  the ground-state energy
with $300$ digits in Table~\ref{conv}. We have checked the rate of
convergence with a reference function
of the type used in ref.\cite{wang1-al}. Unfortunately, however, our
study reveals that this reference-function procedure does not accelerate
the rate of convergence, and that it makes the determination of
the energy eigenvalues less reliable.
Table~\ref{tabev1} shows the calculated energies of the ground and
first excited states for the bounded  sextic, octic, dectic and doudectic
anharmonic potentials, in a comparison with the values reported in
\cite{wang1-al} for the unbounded case.
\begin{table}[htbp]
\caption{\footnotesize The calculated energies for the ground and
first excited state for the
bounded quartic anharmonic (potential $V(x)=x^2+x^4$), where
$(2L)$ is the width of the well and
$I$ is the number of the nonvanishing terms in the truncated series of
the wave function.}
\begin{tabular}{cccl}
\hline
$I$ & $L$ & $n$ & $E_n$  \\
\hline
$15$&$1$& $0$ & $2.6365802$\\
&  &  $1$ &    $10.2632444$\\
$25$&$2$& $0$ & $1.39783341$\\
 & & $1$ &      $4.58734092$   \\
$50$&$3$& $0$ &$1.3923516$\\
& & $1$ &      $4.648812$  \\
$125$ &$4$& $0$ & $1.392351641530291855$\\
 & & $1$ &        $4.64881270421207753$\\
$250$ & $5$ & $0$ & $1.392351641530291855657507876609934184$ \\
  & & $1$         & $4.6488127042120775363770329172605844$ \\
$2250$ &$11$ & $0$ & $1.3923516415\ 3029185565\ 7507876609\
9341846000\ 6671122083\ 4088906349 $ \\
&&& $\ \ \, 3238775674\ 3187564652\ 8590973563\ 4677917591\ 2115137534\ 1738817445$ \\
&&& $\ \ \, 5516240463\ 8371304381\ 7869737001\ 3460935168\ 1548420857\ 4889656901$ \\
&&& $\ \ \, 8003055412\ 3664874321\ 8953435715\ 4174093826\ 2405722951\ 9998568711$ \\
&&& $\ \ \, 1814096892\ 2702273638\ 1698111126\ 0310703429\ 3861341959\ 6456848591$ \\
&&& $\ \ \, 8291463489\ 8518858148\ 6302546939\ 214522103$ \\
\hline
\end{tabular}
\label{conv}
\end{table}

\begin{figure}[htbp]
\caption{\footnotesize Dependence of  the ground state energy
of the bounded quartic anharmonic oscillator
(potential $V(x)=x^2+g\,x^4$) on the coupling constant.}
\epsfxsize=7cm
\centerline{\epsfbox{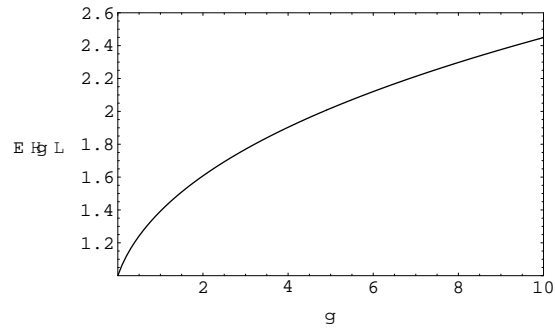}}
\label{valg}
\end{figure}

Figure~\ref{valg} shows the dependence of the ground state energy for
the bounded quartic anharmonic oscillator $(V(x)=x^2+g\,x^4)$ on
the coupling constant $g$,
for $0~\le~g~ \le~10$, and $L=5$. It is clear that the ground-state
energy increases, as expected, when the coupling constant is
increased.

\begin{figure}[hbtp]
\caption{\footnotesize The wave functions for the
ground (left) and first excited (right)
for selected values of the coupling constant $g$,
for the bounded quartic anharmonic oscillator
(potential $V(x)=x^2+g\,x^4$).}
\centering
\begin{minipage}[c]{0.5\textwidth}
\epsfxsize=7cm
\centerline{\epsfbox{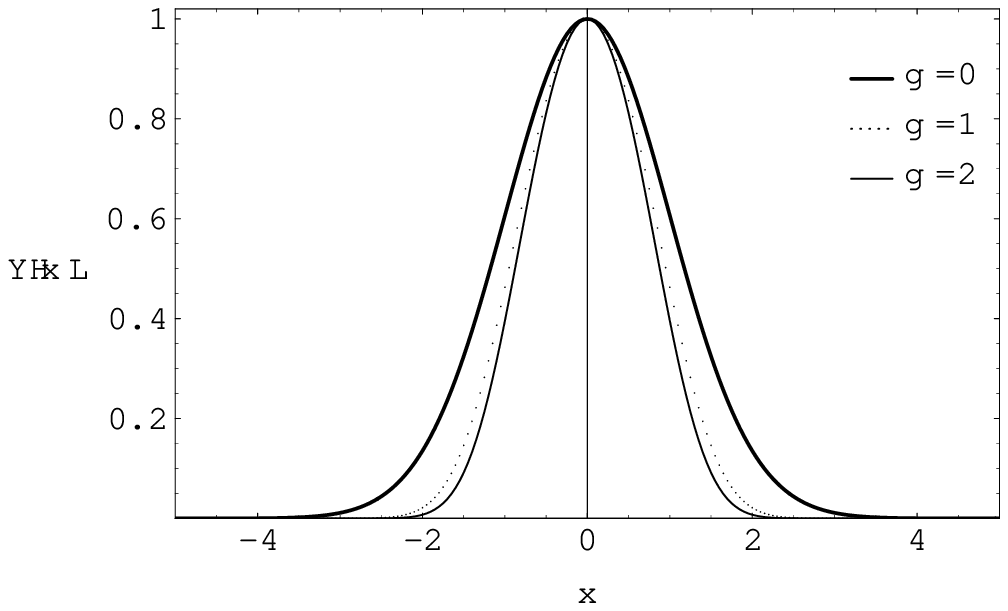}}
\end{minipage}%
\begin{minipage}[c]{0.5\textwidth}
\epsfxsize=7cm
\centerline{\epsfbox{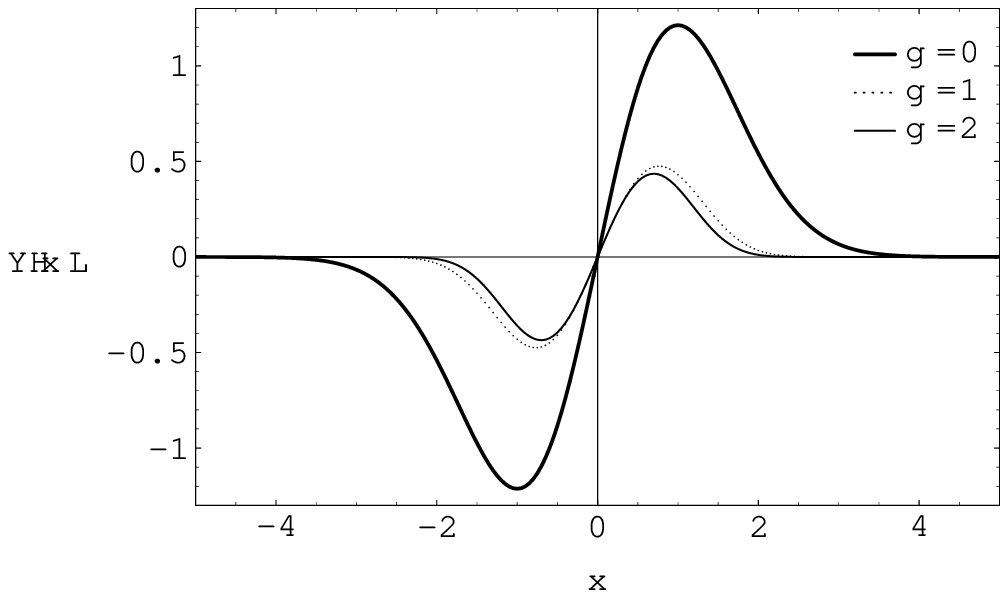}}
\end{minipage}
\label{figqgreg}
\end{figure}

\begin{figure}[hbtp]
\caption{\footnotesize The solid curve represent the ground (left)
and first excited (right) state wave function for the
bounded sextic  anharmonic oscillator $(V(x)=x^2+x^6)$.
The dotted curves represent respectively (inward) the ground
state wave functions for the bounded octic, dectic and duodectic case
$(V(x)=x^2+x^{2k},\ k=4,5,6)$.}
\centering
\begin{minipage}[c]{0.5\textwidth}
\epsfxsize=7cm
\centerline{\epsfbox{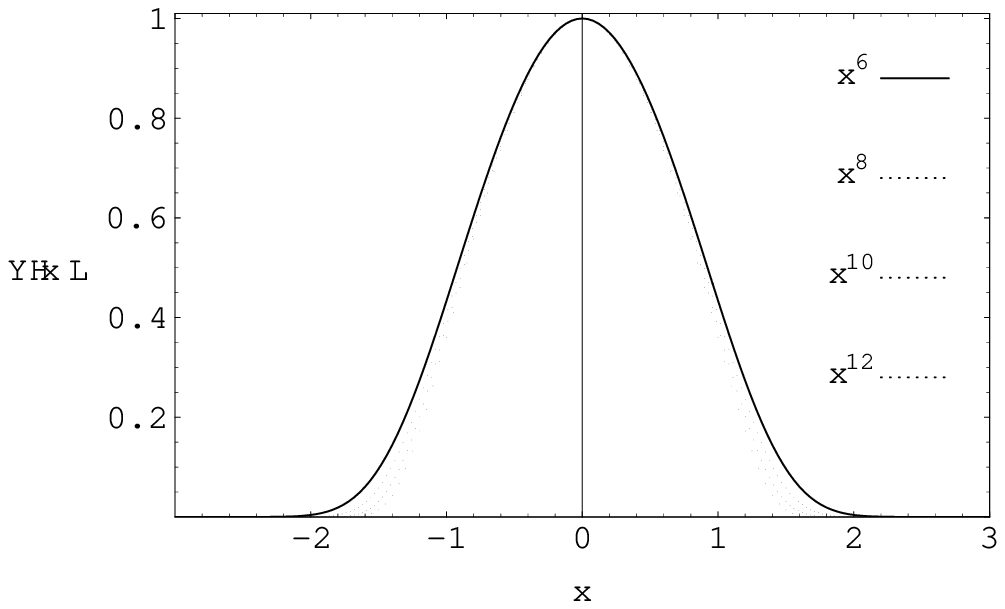}}
\end{minipage}%
\begin{minipage}[c]{0.5\textwidth}
\epsfxsize=7cm
\centerline{\epsfbox{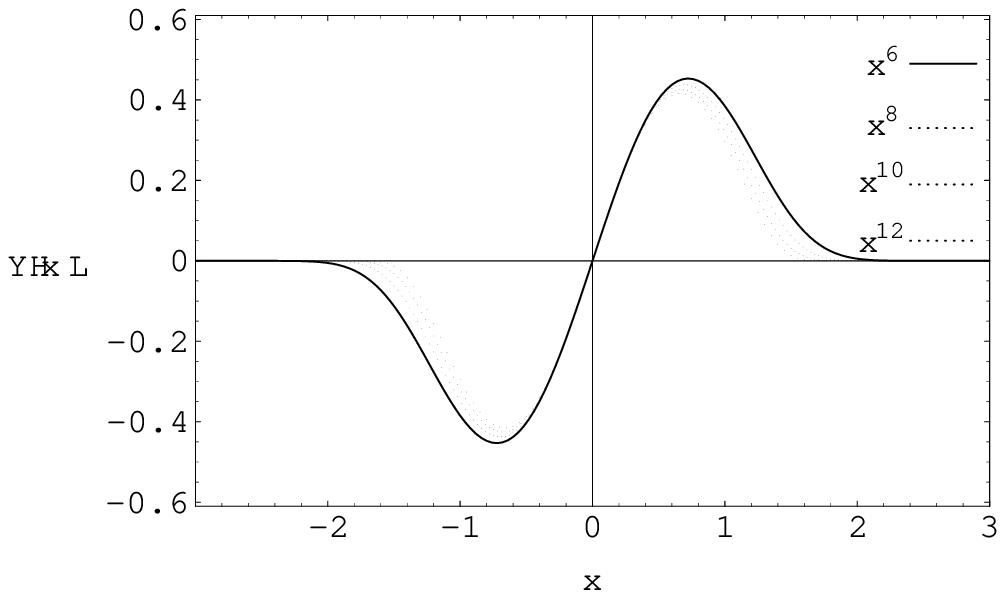}}
\end{minipage}
\label{figsixdue01}
\end{figure}

Figure~\ref{figqgreg} shows the wave functions for the ground state  and
first excited state in the case of the quartic anharmonic oscillator
$(V(x)=x^2+g\,x^4)$ for $g=0, 1, 2$.
Figure~\ref{figsixdue01} shows the ground and
first excited state for the sextic, octic, dectic and duodectic
anharmonic potentials $V(x)=x^2+x^{2k}$ (where $k=3,4,5,6,$ respectively).

\begin{table}[tbp]
\caption{{\footnotesize The calculated energies for the
ground and first excited state for the bounded
quartic, sextic, octic, dectic and duodectic  anharmonic oscillators
(with the indicated  potentials $V(x)$),
where $2 L$ is the width of the well, and where  the underlined
values are as calculated for the unbounded
potential in~\cite{wang1-al}.}}
{\centering
\begin{tabular}{ccl}
\hline
$V(x)$ & $n$ & $E_n$ \\
\hline
\mc{3}{c}{$L=4$}\\
$x^2+x^6$&$0$&
      $\underline{1.435624619003392315}76127222054252$\\
        &$1$&    $5.03339593772026647682838545349365$   \\
\mc{3}{c}{$L=3$}\\
$x^2+x^8$&$0$&
      $\underline{1.491019895662}20496417108006064743$\\
&$1$&            $5.36877806174812976635097601368018$  \\
$x^2+x^{10}$ &$0$&
      $\underline{1.5462635125}7234572711783303167771$\\
&$1$&            $5.65933772479004484069656154796369$\\
\mc{3}{c}{$L=2$}\\
$x^2+x^{12}$ & $0$ &
                  1.59799049927600 \\
& $1$ &           5.91264617503482 \\
\hline
\end{tabular}
}
\label{tabev1}
\end{table}

As emphasized in Sect.~2, all these graphs of the wave
functions show the convergence of the power-series expansion in
the finite interval.

Table~\ref{tabev2} presents the calculated energies of the ground and
first excited states for the bounded double-well  potential,
in comparison with the values reported in \cite{wang1-al}
for the unbounded case.  Our calculations
confirm the predictions made in \cite{wang1-al} concerning
the pseudo-degenerate nature of the ground-state and first-excited-state
energies for $\mu^2=25$, which becomes apparent only after $26$
significant digits.
In addition, we present the energies of the ground state and first excited
state for $\mu^2=35,50$. The splitting for $\mu^2=35$ and $\mu^2=50$
are revealed after $42$ and $72$ significant
digits, respectively.
\begin{table}[tbp]
\caption{{\footnotesize The calculated energies for the
ground and first excited state for the bounded
potential $V(x)=-\mu^2 x^2 +x^4$, where $\mu$ is the mass parameter and
$2 L$ is the width of the well, and where the underlined values are as calculated for the unbounded
potential in ~\cite{wang1-al}.}}
{\centering
\begin{tabular}{ccl}
\hline
$\mu^2$ & Parity & $E_{\pm}$ \\
\hline
\mc{3}{c}{$L=6$} \\
$0$& $+$&$   \hspace{.3cm}\underline{1.06036209\ 04841828\ 99647046\ 016}69266\ 35455152\ 08728529\ $   \\
$ $& $-$&$   \hspace{.3cm}\underline{3.79967302\ 98013941\ 68783094\ 188}51256\ 89577660\ 65467327\ $  \\
$1$& $+$&$   \hspace{.3cm}\underline{0.65765300\ 51807151\ 23059021\ 723}11059\ 35603749\ 37941936\ $    \\
$ $ &$-$&$   \hspace{.3cm}\underline{2.83453620\ 21193042\ 14654676\ 208}74896\ 49582169\ 40153453\ $ \\
$5$ &$+$&$  \underline{-3.41014276\ 12398294\ 75297709\ 653}52190\ 91987123\ 39047565\ $ \\
$$  &$-$&$  \underline{-3.25067536\ 22892359\ 80228513\ 775}54773\ 68771546\ 01147639\ $ \\
\mc{3}{c}{$L=8$}\\
$10$&$+$&$  \underline{-20.63357670\ 29477991\ 49958554}\ 83743150\ 87653159\ 46057736\ $\\
$  $&$-$& $ \underline{-20.63354688\ 44049110\ 79343874}\ 10046139\ 03678429\ 34101495\ $\\
$15$&$+$&$ \underline{-50.84138728\ 43819543\ 66250996\ 515}74123\ 37747896\ 27482985\ $\\
$$& $-$&$  \underline{-50.84138728\ 41870051\ 54710149\ 735}64863\ 44459057\ 68683578\ $\\
$25$&$+$&$ \underline{-149.21945614\ 21908880\ 29163966\ 538}16577\ 44754406\ 92275913\ $\\
$ $&$-$ &$ \underline{-149.21945614\ 21908880\ 29163958\ 974}35901\ 91957349\ 04923409\ $\\
$35$&$+$&$-297.91219449\ 37076184\ 50107115\ 88417483\ 94538893\ 52429871\ $\\
$ $&$-$& $-297.91219449\ 37076184\ 50107115\ 88417483\ 94538891\ 68865338\ $\\
\mc{3}{c}{$L=9$}\\
$50$&$+$&       $-615.02009090\ 27578165\ 66217383\ 21036156\ 72635810\ 55195838$\\
& & $\hspace{1.1cm}06693413\ 44260555\ 97217697\ 29429892\ $\\
$$&$-$&         $-615.02009090\ 27578165\ 66217383\ 21036156\ 72635810\ 55195838$\\
& & $\hspace{1.1cm}06693413\ 44260555\ 97217403\ 20140809\ $\\
\hline
\end{tabular}
}
\label{tabev2}
\end{table}

Our calculations confirm the general qualitative results that the energies
of the low-lying  states of the double-well potential become almost
degenerate (pseudo-degenerate) as the depth of the well increases.

\begin{figure}[hbtp]
\caption{{\footnotesize The wave function for  ground (left)
and first excited (right) state
for the bounded Morse potential,  with depth parameter $V_0= 400$ and range parameter $\lambda=1$,
 where the results for the bounded case are obtained for $I$ (the number
 of nonvanishing terms in the truncated series of the wave function)
 set to $500$  and $L$ (the half-width of the well) set to $2$}}
\centering
\begin{minipage}[c]{0.5\textwidth}
\epsfxsize=7cm
\centerline{\epsfbox{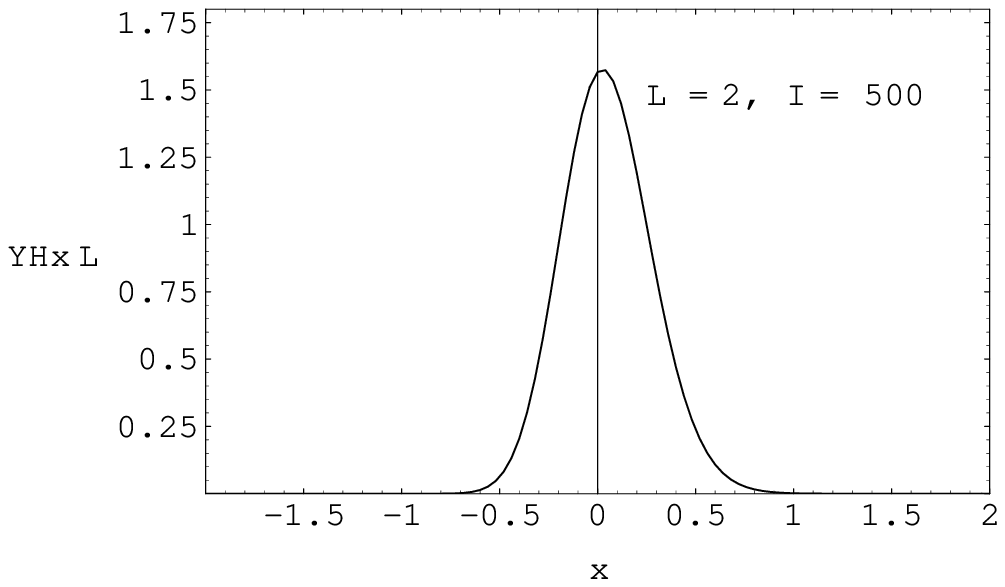}}
\end{minipage}%
\begin{minipage}[c]{0.5\textwidth}
\epsfxsize=7cm
\centerline{\epsfbox{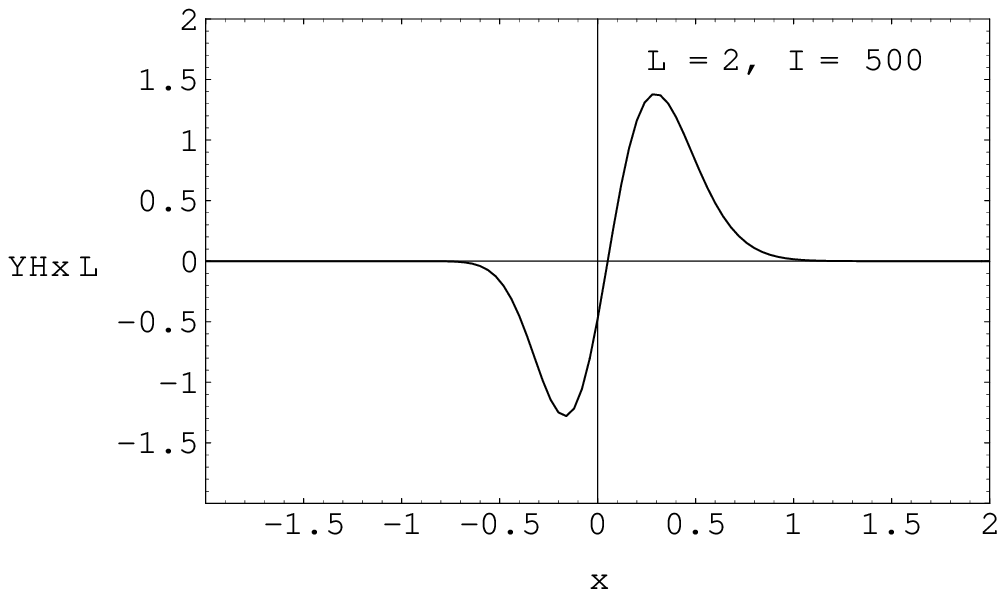}}
\end{minipage}
\label{morsege}
\end{figure}

In the case of the Morse potential with the particular
values $\lambda=1$ and $V_0=400$, we obtain  the
energy eigenvalues tabulated in Table~\ref{resmorse}. Our values
are identical with the exact values to the order indicated in
the table. Figure.~\ref{morsege}
shows the ground state and the first excited state, respectively, for
a bounded Morse
potential confined to the interval $ -2  \le x \le 2$. These
figures match the exact corresponding wave function of the
unbounded Morse potential in the same interval.
\begin{table}[htbp]
\caption{{\footnotesize Energy eigenvalues for the Morse potential
with depth parameter $V_0= 400$ and range parameter $\lambda=1$,
 where the results for the bounded case are obtained for $I$ (the number
 of nonvanishing terms in the truncated series of the wave function)
 set to $500$  and $L$ (the half-width of the well) set to $2$}.}
{\centering
\begin{tabular}{ccccc}
\hline
 & $E_0$ & $E_1$ & $E_2$ & $E_3$  \\
\hline
& $19.75000000000$&$57.75000000000$& $93.75000000000$ & $127.75000000$\\
Exact& $19.75$ &  $57.75$ & $93.75$ & $127.75$ \\
\hline
\end{tabular}
}
\label{resmorse}
\end{table}
\section{Conclusion}
In this paper, we have applied the method of power-series expansion to a
variety of symmetric one-dimensional anharmonic potential functions
bounded by two infinite walls. We have compared our calculations
of the low-lying energy levels with results from the inner-product
quantization procedure. For all the potentials considered,
we have chosen
values for the number $n$ of terms in the power series and  the distance
$2\,L$ between the walls that yield results of an accuracy
superior to the results from the inner-product quantization procedure.
More importantly, in analyzing the ground
and first excited states of the deep  double well, we have found the
splitting of the energy levels to become apparent only
after $26$ significant digits for the mass parameter $\mu^2=25$,
in agreement with the
result obtained from the inner-product quantization procedure. We have
extended our calculation to the case  of $\mu^2=35$ and $\mu^2=50$,
where the splitting shows up after $42$  and $72$ significant digits,
respectively. For sextic, octic, dectic and duodectic oscillators,
we have calculated the energy and wave function for
the first excited states as well as for the ground states.
Finally, we have applied our method to
the well-known Morse potential, finding that our numerical results
match the exact results.

A reason for the capability of the present method
is that for a bound state, the wave function is spatially localized. This
means that the probability density $|\Psi|^2$ has appreciable values in a
finite region of space, outside which the probability density tends
rapidly to zero. To obtain a good approximation, it is thus reasonable,
as is shown by our numerical results for our chosen example of
the Morse potential, to consider the corresponding problem in a finite
interval, with a suitable width, bounded by two infinite walls.

Although the two methods
are equally powerful for the potentials considered in the present work,
the power-series expansion in the finite  range is
convergent for potentials admitting a power-series expansion in
the same finite interval. The power-series method has been also applied
to the case of one dimensional multi-well oscillator \cite{ourmw}.
In addition, it has been used in~\cite{zinconj} to
justify numerical results based on the Zinn-Justin conjecture
\cite{zinnlett}.
We show in work under preparation that the power-series
method can be extended to the case of three-dimensional spherically
symmetric potentials.

\section*{Acknowledgement}
This work was supported  by the Research Center, College of Science,
King Saud University under project number Phys$/1423/02$.


\begin{thebibliography}{99}
{\small
\bibitem{landau} L. D. Landau and E. M. Lifshitz, {\it Quantum mechanics
(Pergamon Press, Oxford, 1977), 3rd ed.}
\bibitem{coleman-dw} S. Coleman, {\it In The Whys of Subnuclear
Physics}, edited by A. Zichichi (Plenum, New York, 1979), pp.
270-274.
\bibitem{barakat-an} R. Barakat and R. Rosner, \pla{83}{149}{81}.
\bibitem{chaud-an} R. N. Chaudhuri and B. Mukherjee,
\jpa{16}{3193}{1983};\ibid{17}{3327}{1984}.
\bibitem{haut-an} A. Hautot, \prd{33}{437}{86}.
\bibitem{voros-an} A. Voros, \jpa{27}{4653}{1994}; \ibid{32}{5993}{1999}.
\bibitem{trot-an}  M. Trott {\it quant-ph/0012147}\,{\bf (2000)}.
\bibitem{wang1-al} C. J. Tymczak, G. S. Japaridze, C. R. Handy, and X.
-Q. Wang, \prl{80}{3673}{98}.
\bibitem{cheb-al} I. V. Chebotarev, \ap{273}{114}{99}.
\bibitem{roy-al} A. K. Roy, N. Gupta, and B. M. Deb,
\pra{65}{012109}{2002}.
\bibitem{codd} E. A. Coddington,  An introduction to ordinary
differential equations,
(Prentice-Hall,Inc., Englewood Cliffs,  N.J., 1966), pp. 138-142.
\bibitem{ourmw} H. A. Alhendi and E. I. Lashin {\it
quant-ph/0306016}\,{\bf (2003)}.
\bibitem{zinconj}H. A. Alhendi and E. I. Lashin, \jpa{38}{6785}{2005}, {\it
quant-ph/0402101}\, {\bf (2004)}.
\bibitem{zinnlett} U. D. Jentschura and J. Zinn-Justin,
\jpa{34}{L253 }{2001}.
}
\end{thebibliography}
\end{document}